# Optoelectronic and transport properties of Fe$_2$TiGe with the Tran-Blaha modified Becke-Johnson potential

M. Anwar Hossain*, M. A. Al Noman and Md. Taslimur Rahman

Department of Physics, Mawlana Bhashani Science and Technology University, Santosh, Tangail-1902, Bangladesh
Email*: anwar647@mbstu.ac.bd

Abstract

In this paper, we have performed first principles calculations to study optoelectronic, thermodynamic and transport properties of Fe$_2$TiGe using density functional theory (DFT). The semi-classical Boltzmann transport theory is used to investigate transport properties. The calculated energy bands indicate that Fe$_2$TiGe is an indirect band gap semiconductor with band gap 0.734 eV for TB-mBJ functional. Fe-3d and Ti-3d orbitals have the dominant contributions to the density of states due to strong hybridization between them. The maximum value of absorption coefficient is found to be $224 \times 10^4$ cm$^{-1}$ in the ultraviolet region. It is found that the static refractive index of Fe$_2$TiGe is 5.18 which is very close to that of Ge (5.974) and much higher than that of GaAs (3.29–3.857). The obtained refractive index implies that Fe$_2$TiGe is a potential optical material that can be used in optical devices such as photonic crystal, wave guides and solar cells. The Debye temperature at ambient condition is 570 K and decreases slowly with temperature. The Grüneisen parameter at ambient pressure and 300 K is ~2.1 and slightly higher than that of PbTe (1.5). The calculated Seebeck coefficient and power factor at 300 K using TB-mBJ functional are 271 µV/K and 5.9 µWcm$^{-1}$K$^{-2}$, respectively. Therefore, the calculated optoelectronic properties implies that Fe$_2$TiGe is a potential candidate for photovoltaic device applications.

1. **Introduction**

Energy is the key ingredient for all kinds of development and its necessity is increasing rapidly to hold the sustainability of development. The natural sources of energy are fossil fuels like gas, coal and oil and these are non-renewable. Fossil Fuels are rapidly diminishing and hence a huge attention has been paid to find the alternative energy sources and energy efficient materials. Solar energy can help us greatly to face the biggest challenges of energy crisis in the 21$^{st}$ century. The materials with high absorbance and tunable band gap in the visible range of light are suitable for photovoltaic or solar energy applications. On the other hand, one of the hot topics of energy research is the development of efficient thermoelectric materials that are used to convert waste heat (generating in many human activities such as thermal power plants, cement plants, steelworks, vehicles engines, factories, incinerators, etc) to electricity. The efficiency of a potential thermoelectric material is determined by the dimensionless figure of merit, $ZT=\frac{S^2\sigma T}{k}$, where $S$, $\sigma$, $k = k_e + k_l$ and $T$ are the Seebeck coefficient, electrical conductivity, thermal conductivity (consisting of electronic and lattice thermal conductivity) and absolute temperature, respectively. To realize the efficient thermoelectric energy conversion, the thermoelectric materials should have low thermal conductivity ($\kappa$), high electrical conductivity ($\sigma$), and large Seebeck coefficient ($S$). In the last decade, the intermetallic Heusler compounds has gained tremendous attention because of their interesting physical properties such as half-metallicity, spin gapless semiconductivity, giant magnetocaloricity, thermoelectricity and superconductivity [1–6]. In various technological applications these properties have been utilized [7–11]. The general formula of Fe$_2$ based Heusler compounds is Fe$_2YZ$ ($Y$ is transition metal and $Z$ is a main group element) and these compounds exhibit very interesting structure, electronic and magnetic properties [12–14]. Fe$_2$YZ display a large magnetization and high Curie temperature are of special interest for potential magnetic applications [15,16]. V. Sharma and G. Pilania studied the electronic, magnetic, optical and elastic properties of Fe$_2$YAl (Y=Ti, V and Cr ) and they obtained 100% spin polarization in Fe$_2$CrAl compound in the vicinity of Fermi level [12]. The half-metallic ferromagnets (HMFs) have intensively studied for their potential applications in the field of spintronics [17–19]. Z. Ren et *al.* studied structure and magnetic properties of Fe$_2$CoGe synthesized by ball-milling and the ferromagnetic Fe$_2$CoGe does not show half metallic character [20]. It has been reported that Fe$_2$VAl-based full-Heusler alloys show a large power factor which is substantially higher than that of Bi$_2$Te$_3$ [21] and this higher power factor implies that full-Heusler alloys are potential candidate for thermoelectric applications. The total number of valence electrons in Fe$_2$VAl and Fe$_2$TiSn are 24 and according to Slater–Pauling rule the total spin magnetic moment vanishes and these two alloys show semi metallic or semiconducting properties [1, 22–24]. The thermoelectric properties of Fe$_2$VAl has been studied intensively [21, 25–29] and increasing attention toward the search for and study of new thermoelectric materials based on full-Heusler alloys has been observed over the last few years [30, 31]. Hongzhi Luo et al. reported that Fe$_2$TiGe and Fe$_2$TiSn are nonmagnetic semiconductors [32]. The origin of large Seebeck coefficients in semiconducting Fe$_2$TiSn and Fe$_2$TiSi have been investigated by Shin Yabuuchi et al. [33]. Both Fe$_2$TiSi and Fe$_2$TiSn have flat bands at the bottom of the conduction band along Γ–X directions and possess a high Seebeck coefficient [33]. Recently Bhat *et*.al studied the thermoelectric properties of Fe$_2$TiGe, Fe$_2$TiSi and Fe$_2$ZrSi using PBE potential and they found semiconducting nature in electrical conductivity and Seebeck coefficient [34]. The high value of Seebeck coefficient and low resistivity is the prerequisite for promising thermoelectric materials. The TB-mBJ functional yields band gaps in good agreement with experiment in contrast to the local density approximation (LDA) or the Perdew, Burke and Ernzerhof (PBE) functionals. Therefore TB-mBJ potential is suitable to predict the accurate band gap and a lot of theoretical studies have been carried out to explore thermoelectric transport properties using this potential [35–41]. The precise band gap calculation is essential to study the optical properties of semiconducting materials. The optical and thermodynamic properties of $Fe_2TiGe$ is still unexplored.

The above discussion motivate us to study the optoelectronic, thermodynamic and thermoelectric transport properties of nonmagnetic semiconductor $Fe_2TiGe$ using TB-mBJ exchange potential. In this paper, we have presented first-principles study of electronic, optical, thermodynamic and thermoelectric transport properties of $Fe_2TiGe$ by using density functional theory (DFT) [42-43] and semi-classical Boltzmann transport theory [44]. The calculated band structure and density of states of $Fe_2TiGe$ confirms the semiconducting nature and the predicted optoelectronic properties imply that $Fe_2TiGe$ is a potential material for optoelectronic device applications.

## 2. Computational details

Electronic and optical properties were studied by using the full potential linearized augmented plane wave (LAPW) implemented in WIEN2k [45]. For good convergence, a plane wave cutoff of kinetic energy $RK_{max}$ =8.0 and (15 × 15 × 15) *k*-point in Brillouin zone integration were selected. The muffin tin radii 2.42 for Fe, 2.3 for Ti, and 2.3 for Ge, were used. The PBE and modified TB-mBJ [46] functional were used in the electronic structure, optical properties and transport properties calculations. The convergence criteria of energy and charge were set to $10^{-4} Ry$ and 0.001e, respectively. The transport properties were calculated by BoltzTraP code [47]. For this, we used (43 × 43 × 43) *k*-point in WIEN2k to generate the required input files and the chemical potential was set to the zero temperature Fermi energy. By solving the semi-classical Boltzmann transport equation, we can easily calculate the transport coefficients. The transport coefficients are defined in the Boltzmann transport theory [48–50] as

$$\sigma_{\alpha\beta}(T,\mu) = \frac{1}{V}\int \Sigma_{\alpha\beta}(\varepsilon)\left[-\frac{\partial f_\mu(T,\varepsilon)}{\partial \varepsilon}\right]d\varepsilon \tag{1}$$

$$S_{\alpha\beta}(T,\mu) = \frac{1}{eTV\sigma_{\alpha\beta}}\int \Sigma_{\alpha\beta}(\varepsilon)(\varepsilon-\mu)\left[-\frac{\partial f_\mu(T,\varepsilon)}{\partial \varepsilon}\right]d\varepsilon \tag{2}$$

$$K^e_{\alpha\beta}(T,\mu) = \frac{1}{e^2 TV}\int \Sigma_{\alpha\beta}(\varepsilon)(\varepsilon-\mu)^2\left[-\frac{\partial f_\mu(T,\varepsilon)}{\partial \varepsilon}\right]d\varepsilon \tag{3}$$

where V is the volume of a unit cell, $\alpha$ and $\beta$ represent Cartesian indices, $\mu$ is the chemical potential and $f_\mu$ Fermi-Dirac distribution function. The energy projected conductivity tensors can be calculated by using the following equation

$$\Sigma_{\alpha\beta}(\varepsilon) = \frac{e^2}{N}\sum_{i,k}\tau_{i,k}v_\alpha(i,k)v_\beta(i,k)\frac{\delta(\varepsilon-\varepsilon_{i,k})}{d\varepsilon} \tag{4}$$

where N is the number of k-points for BZ integration, $i$ is the index of band, $v$ and $\tau$ represent the electrons group velocity and relaxation time, respectively. In BoltzTraP program, the constant relaxation time approximation (CRTA) is used.

## 3.    Result and Discussions

The equilibrium crystal structure of Fe$_2$TiGe is shown in Fig.1. The Fe$_2$TiGe is a face-centered cubic crystal with space group $Fm\bar{3}m$ (225). The occupied Wyckoff position for Fe, Ti and Ge atoms are 8c (0.25, 0.25, 0.25), 4b (0.5, 0.5, 0.5) and 4a (0, 0, 0), respectively [51]. The optimized lattice parameter is 5.79 Å and this is very close to theoretically predicted value (5.74Å) [34].

## 3.1. Electronic properties

The electronic band structures of $Fe_2TiGe$ are presented in Fig.2. The bands are non-dispersive and there are no overlapping between the valence band and conduction band at the Fermi level. The conduction band minimum (CBM) and valence band maximum (VBM) are found at X and Γ-points, respectively. Thus, $Fe_2TiGe$ is an indirect band gap semiconductor. It is found that the band gap obtained using PBE-GGA potential is much smaller than that obtained by TB-mBJ potential. Our calculated band structure of $Fe_2TiGe$ using PBE-GGA potential is consistent with others previous result [52]. The values of band gaps obtained from PBE-GGA and TB-mBJ exchange potentials are 0.145 eV and 0.734 eV, respectively. The band gap obtained from TB-mBJ potential is much closer to that of TiPdSn (0.74 eV) [53] The energy bands of $Fe_2TiGe$ are mainly d-like character, although Ge-p orbitals have very little contributions as indicated in Fig. 3. From the band structure (Fig. 2), one can observe a flat band in the CB along the Γ to X point of the Brillouin zone. It is also observed that a narrow Lorentzian peak in the VBM at the Γ point. These structural properties indicate that $Fe_2TiGe$ is potential thermoelectric materials. The total and atomic density of states are shown in Fig. 3. The upper peak in the density of states, from -3 to -1eV, comes from the strong hybridization between Fe-3d and Ti-3d orbitals. This can be well explained by sigma bonding combinations of Fe-3d and Ti-3d orbitals. The lower peak, from 1 to 3 eV, arises from the corresponding anti-bonding combinations. The calculated band structures and density of states of $Fe_2TiGe$ using PBE-GGA and TB-mBJ potentials are consistent.

## 3.2. Optical properties

The optical properties of a material are entirely depend on its electronic band structure. A material suitable for optoelectronic or photovoltaic application must have high absorbance, high refractive index and low emissivity of light. Moreover high dielectric constant, high optical conductivity and reflectivity are prerequisites for a good optical materials. For certain photon energies the optical response can be described by frequency dependent complex dielectric function $\varepsilon(\omega)$ expressed by Ehrenreich and Cohen as $\varepsilon(\omega) = \varepsilon_1(\omega) + i\varepsilon_2(\omega)$ [54], where $\varepsilon_1(\omega)$ and $\varepsilon_2(\omega)$ are the real and imaginary part of dielectric function. The real part is related to the polarization, anomalous dispersion and the imaginary part is related to the loss of energy into the medium. These two parts of dielectric function completely explain the optical properties of materials at all photon energies [55]. The transition between the valence and conduction bands determines the imaginary part, $\varepsilon_2(\omega)$ and the real part, $\varepsilon_1(\omega)$ of the dielectric constant can be calculated using the Kramers-Kronig dispersion relation [56, 57]. The inter-band transitions in semiconductor have major contribution to the dielectric function. Here we have studied the optical properties of $Fe_2TiGe$ in terms of its corresponding parameters in the inter-band region. The calculated optical parameters such as real part $\varepsilon_1(\omega)$, imaginary part $\varepsilon_2(\omega)$ of complex dielectric function $\varepsilon(\omega)$, refractive index $n(\omega)$, extinction coefficient $k(\omega)$, absorption coefficient $\alpha(\omega)$, optical conductivity $\sigma(\omega)$, energy loss function $L(\omega)$ and reflectivity $R(\omega)$ versus photon energy up to 10 eV are presented in Fig. 4 (a-h). The real part of dielectric function $\varepsilon_1(\omega)$ is maximum at zero photon energy of incident photon shown in Fig. 4 (a) and the values are found to be 37.43 and 26.86 for PBE and TB-mBJ functional, respectively. These values of $\varepsilon_1(0)$ are greater than the values of well know photovoltaic materials silicon (11.7) and GaAs (12.9), respectively. The spectra of the imaginary part of dielectric function $\varepsilon_2(\omega)$ is completely opposite in fashion than that of $\varepsilon_1(\omega)$ and directly connected with the energy band structure shown in Fig. 4 (b). The critical points of $\varepsilon_2(\omega)$ are located around 0.15 and 0.70 eV for PBE and TB-mBJ functional, respectively, known as fundamental absorption edge. These values of fundamental absorption edge are in good agreement with our calculated energy band gap for both functionals. The major peaks correspond to $\varepsilon_2(\omega)$ are found around 2 and 4 eV due to the first order transition between topmost valence band of Fe-3d state and the lowermost conduction band of Ti-3d state where breaking of the full translational symmetry is responsible for this. Other two peaks are originated by interband transition from Ge-4p valance state to Fe-3d and Ti-3d conduction states. The calculated refractive index n(ω) for different functional are illustrated in Fig. 4 (c). It is found that the static refractive index of $Fe_2TiGe$ is 5.18 which is very close to that of Ge (5.974) [58–61] and much higher than that of GaAs (3.29–3.857) [62, 63]. The obtained refractive index implies that $Fe_2TiGe$ is a potential optical material that can be used in optical devices such as photonic crystal, wave guides and solar cells. The extinction coefficient k (ω) is illustrated in Fig. 4 (d), where the significant peak is found near UV region and decreases with increase in photon energy. It is found from Fig. 4 (e) that the absorption coefficient α (ω) of $Fe_2TiGe$ starts rising at around 0.15 and 0.70 eV for PBE and TB-mBJ functionals, respectively which is consistent with the predicted band gap. The maximum absorption coefficients for TB-mBJ potential is found to be $224\times10^4$ $cm^{-1}$ which is larger than that of GaAs ($(14–22)\times10^4$ $cm^{-1}$ at ~4.8 eV) [64, 65] and silicon ($1.8\times10^6$ $cm^{-1}$) [66]. In Fig. 4 (f), the optical conductivity σ (ω) does not start at zero photon energy since the studied material ($Fe_2TiGe$) has a distinct band gap but starts to rise when the incident photon energy is higher than the band gap which is also an evident from the calculated band structure. Moreover, the optical conductivity and hence electrical conductivity of a material increases due to photon absorption [67]. The maximum optical conductivity ($14\times10^3$ S/cm) for TB-mBJ potential is obtained around 3 eV of incident photon energy. The energy loss function L(ω) and the reflectivity R(ω) are presented Fig. 4 (g) and (h), respectively. The energy loss of a fast electron through $Fe_2TiGe$ is very low and maximum energy loss occurs at 9.26 eV. The most prominent peaks correspond to the plasmon peaks and the frequency (energy) at which it occurs is known as plasma frequency which indicate the transition from metallic to dielectric nature. The calculated zero frequency reflectivity using TB-mBJ is 0.46. The reflectivity for TB-mBJ functional is much lower in the visible and ultraviolet region which makes sure its potential applications in the area of transparent coatings [68].

### 3.3. Thermodynamic properties

The thermodynamic properties are related to the stability and durability of a crystal. The thermal properties such as specific heat and Debye temperature are related with thermal conductivity. A good thermoelectric material requires as much as low thermal conductivity to convert larger amount of waste heat to electricity. The Debye temperature decreases with increasing temperature for intermetallic compounds [69]. The materials with low Debye temperature can possess low lattice thermal conductivity. So it is interesting to investigate Debye temperature of $Fe_2TiGe$. The equilibrium energy, bulk modulus, volume, and other required parameters to calculate thermodynamic quantities were obtained by volume optimization. The Murnaghan equation of state was used to fit the energy versus volume and thus using standard thermodynamic relations the macroscopic thermodynamic quantities as a function of pressure (P) and temperature (T) were calculated. The temperature and pressure dependent thermodynamic parameters are illustrated in Fig. 5. Fig. 5(a) and 5(b) illustrates the variations of bulk modulus and Debye temperature with temperature at different pressure for $Fe_2TiGe$ and it is found that bulk modulus and Debye temperature increases rapidly with the increase in pressure but slowly decreases with temperature. Since Debye temperature is inversely related to the vibrational frequency, these results show that vibrational frequency increases with the increase in temperature but decreases due to the increase in pressure. Thus, phonon contribution increases with the increase in pressure but decreases with increase in temperature. The bulk modulus determine the degree of resistance of a material. A material will be more resistive when the bulk modulus is larger. Since the pressure reduces the lattice parameters, therefore, bulk modulus is inversely related to lattice parameters. Thus, the degree of resistance of $Fe_2TiGe$ decreases with increasing temperature but increases with increasing pressure. This behavior is responsible for the change in volume due to pressure and temperature effect. It is clear from Fig. 5(c) that there is no significant pressure effect on $C_V$ within the studied temperature range. As temperature increases, $C_V$ approaches the Dulong-Petit limit, as expected for any solids [70]. It is found that at 0 GPa and 300K, the value of $C_V$ is approximately 83.85 J/mol.K. The Grüneisen parameter determines the anharmonicity and hence phonon scattering. The small Grüneisen parameter indicates the small phonon scattering which may lead to large contributions to the lattice thermal conductivity. The Grüneisen parameter decreases with pressure but increases slowly with temperature as expected.

### 3.4. Thermoelectric transport properties

A good thermoelectric material should be narrow band gap semiconductor with flat bands. Our calculated band structure confirms that $Fe_2TiGe$ is an indirect band gap semiconductor with flat conduction band directed from Γ- to X k-point. Furthermore, the density of states confirms the bulk gap between valence and conduction bands. Recently, Bhat et al. predicted thermoelectric transport in $Fe_2TiGe$ using PBE-GGA potential [34]. Therefore, it is interesting to study the thermoelectric transport properties of $Fe_2TiGe$by considering TB-mBJ density functional. We have calculated thermoelectric parameters as a function of chemical potential and temperature within constant relaxation time approximation (cRTA) are presented in Fig. (6, 7). Thermoelectric transport parameters around the Fermi level for PBE-GGA (left panel) and TB-mBJ functionals

(right panel) at 300, 500 and 800 K are shown in Fig. 6(a-j) where the significant values of these parameters are found in the range of chemical potential of -0.05 to 0.05 eV. The calculated Seebeck coefficient considering TB-mBJ functional is larger than that of PBE-GGA functional since the band gap for TB-mBJ functional is higher than PBE-GGA functional. The maximum values of Seebeck coefficient for TB-mBJ functional around the Fermi level at 300, 500 and 800 K are 1086, 665 and 416 µV/K, respectively. The calculated electrical conductivity and electronic thermal conductivity are minimum at zero chemical potential for both functional as shown in Fig. 6(b, g) and (c, h). The electrical conductivity around the Fermi level is almost independent of temperature but electronic thermal conductivity increases with the increase of temperature. This increase in electronic thermal conductivity is expected because carrier concentration increases with the increase of temperature in semiconducting materials.

The power factor of conduction band is larger than that of valence band which is shown in Fig. 6(d, i). Using TB-mBJ potential the maximum power factor ($S^2\sigma/\tau$) predicted at 300, 500 and 800 K from 0.01 to 0.03 eV in the conduction band are 46, 65 and 82 µWcm$^{-1}$K$^{-2}$, respectively. The power factor calculated using TB-mBJ is much higher than that of recent studied compound Fe$_2$TiSn [34]. It is obvious from Fig. 6(e, j) that for TB-mBJ potential the maximum ZT values at 300 in the valence and conduction bands are 0.98 and 0.99, respectively. Fig. 6 shows the temperature dependence of thermoelectric transport parameters of Fe$_2$TiGe. The temperature dependence of Seebeck coefficient (*S*) is presented in Fig. 7 (a). After 300 K, Seebeck coefficient decreases sharply in the case of PBE potential while it is decreases very slowly when the applied potential is TB-mBJ. The Seebeck coefficient obtained at 300 K for PBE and TB-mBJ potentials are 228.7 µV/K and 254 µV/K, respectively which is little smaller than that of TiPdSn [71]. The variation of electrical conductivity with temperature is shown in the Fig. 7 (b). The electrical conductivity (σ/τ) increase rapidly with temperature and this behavior indicates the semiconducting nature of $Fe_2TiGe$. The calculated electrical conductivity (σ/τ) of $Fe_2TiGe$ at 300 K using PBE and TB-mBJ are almost same and the value is $1.0 \times 10^{18}$ 1/Ωms. Power factor after 400 K decreases for PBE potential whereas it is increasing rapidly with temperature for TB-mBJ potential. The maximum power factor (28.5 µWcm$^{-1}$K$^{-2}$ with τ = 10$^{-15}$ s) is obtained at 800 K with TB-mBJ potential. The temperature dependence of electronic thermal conductivity ($\kappa_e$/τ) is illustrated in Fig. 7(c). The electronic thermal conductivity is directly related with electrical conductivity ($k_e=L\sigma T$) and increases with temperature in a very linear fashion as shown in Fig. 7 (d). The low electronic thermal conductivity is obtained from 100 to 300 K and this is due to the low electrical conductivity. For a good thermoelectric material low thermal conductivity and high power factor are essential. The dimensionless figure of merit of $Fe_2TiGe$ has been calculated by plugging the desired quantities in $ZT = \frac{S^2(\sigma/\tau)}{(k_e/\tau)}T$, is depicted in Fig. 7 (e). Here we considered only electronic thermal conductivity to calculate *ZT*. We see that *ZT* decreasing with temperature for both potentials but the decrease rate for PBE potential is much faster that TB-mBJ potential. The calculated *ZT* value at 300 K for PBE and TB-mBJ potentials are 0.68 and 0.80, respectively and this values are very close to that for TiPdSn [71], indicates that $Fe_2TiGe$ could be a high performance thermoelectric material. It is note that the calculated thermoelectric properties using TB-mBJ is more accurate than that of PBE potential as mentioned in introduction.

**4. Conclusions**

In summary, we have performed first principles calculations to study electronic, thermodynamic, optical and thermoelectric properties of $Fe_2TiGe$ using density functional theory (DFT). The semi-classical Boltzmann transport theory is used for transport properties. The calculated energy bands indicate that $Fe_2TiGe$ is an indirect band gap semiconductor and the value of gap are 0.145 eV and 0.734 eV using PBE and TB-mBJ potentials, respectively. Such increase of band gap due to mBJ potential calculation has significant effect on transport properties. The Fe-3d and Ti-3d orbitals have dominant contributions to the density of states due to strong hybridization between them. The bulk modulus is a measure of degree of resistance of materials. The temperature dependence of bulk modulus at different pressure implies that the degree of resistance of $Fe_2TiGe$ decreases with temperature and increases with pressure. The calculated specific heat at constant volume increases with temperature rapidly up to 500 K and then approaches very slowly to a constant value at high temperature following Dulong-Petit limit, 98 J/mol.K. The maximum value of absorption coefficient is found to be $224 \times 10^4$ cm$^{-1}$ and $202 \times 10^4$ cm$^{-1}$ in the ultraviolet region and the static refractive index is found to be very high (5.18) with TB-mBJ density functional. The static refractive index of $Fe_2TiGe$ is 5.18 which is very close to that of Ge (5.974) [58–61] and much higher than that of GaAs (3.29–3.857) [62, 63]. The predicted refractive index implies that $Fe_2TiGe$ is a potential optical material that can be used in optical devices such as photonic crystal, wave guides and solar cells. The calculated Seebeck coefficient using PBE and TB-mBJ potentials at 300 K are 228.7 and 254 μV/K, respectively. The electronic thermal conductivity increases with temperature as the carrier concentration increases. This behavior indicates the semiconducting nature of $Fe_2TiGe$. The predicted ZT value using TB-mBJ potential at 300 K is ~0.8. The obtained Grüneisen parameter indicates the small phonon scattering which may lead to large contributions to the lattice thermal conductivity. Here we did not calculate the lattice thermal conductivity and the reduction of it could be achieved by doping or nanostructuring in Fe$_2$TiGe system. Therefore, the calculated results implies that Fe$_2$TiGe is a potential candidate for photovoltaic device applications.

**References**

[1] I. Galanakis, I. Galanakis, PH Dederichs, and N. Papanikolaou, Phys. Rev. B 66, 174429 (2002)., Phys. Rev. B. 66 (2002) 174429.

[2] J. Pons, E. Cesari, C. Segu'\i, F. Masdeu, R. Santamarta, Ferromagnetic shape memory alloys: alternatives to Ni--Mn--Ga, Mater. Sci. Eng. A. 481 (2008) 57–65.

[3] K. Özdoğan, E. Şaşioğlu, I. Galanakis, Slater-Pauling behavior in LiMgPdSn-type multifunctional quaternary Heusler materials: Half-metallicity, spin-gapless and magnetic semiconductors, J. Appl. Phys. 113 (2013).

[4] A. Planes, L. Mañosa, M. Acet, Magnetocaloric effect and its relation to shape-memory properties in ferromagnetic Heusler alloys, J. Phys. Condens. Matter. 21 (2009) 233201.

[5] D. Do, M.-S. Lee, S.D. Mahanti, Effect of onsite Coulomb repulsion on thermoelectric properties of full-Heusler compounds with pseudogaps, Phys. Rev. B. 84 (2011) 125104.

[6] J. Winterlik, G.H. Fecher, A. Thomas, C. Felser, Superconductivity in palladium-based Heusler compounds, Phys. Rev. B. 79 (2009) 64508.

[7] T. Saito, N. Tezuka, M. Matsuura, S. Sugimoto, Spin injection, transport, and detection at room temperature in a lateral spin transport device with Co2FeAl0. 5Si0. 5/n-GaAs schottky tunnel junctions, Appl. Phys. Express. 6 (2013) 103006.

[8] T. Kubota, S. Tsunegi, M. Oogane, S. Mizukami, T. Miyazaki, H. Naganuma, Y. Ando, Half-metallicity and Gilbert damping constant in Co 2 Fe x Mn 1- x Si Heusler alloys depending on the film composition, Appl. Phys. Lett. 94 (2009) 122504.

[9] Ikhtiar, S. Kasai, A. Itoh, Y.K. Takahashi, T. Ohkubo, S. Mitani, K. Hono, Magneto-transport and microstructure of Co2Fe (Ga0. 5Ge0. 5)/Cu lateral spin valves prepared by

top-down microfabrication process, J. Appl. Phys. 115 (2014) 173912.

[10]  J. Winterlik, S. Chadov, A. Gupta, V. Alijani, T. Gasi, K. Filsinger, B. Balke, G.H. Fecher, C.A. Jenkins, F. Casper, others, Design Scheme of New Tetragonal Heusler Compounds for Spin-Transfer Torque Applications and its Experimental Realization, Adv. Mater. 24 (2012) 6283–6287.

[11]  S. Ouardi, G.H. Fecher, C. Felser, J. Kübler, Realization of spin gapless semiconductors: The Heusler compound $Mn_2CoAl$, Phys. Rev. Lett. 110 (2013) 100401.

[12]  V. Sharma, G. Pilania, Electronic, magnetic, optical and elastic properties of $Fe_2YAl$ (Y= Ti, V and Cr) using first principles methods, J. Magn. Magn. Mater. 339 (2013) 142–150.

[13]  A. Ayuela, A. Ayuela, J. Enkovaara, K. Ullakko, and RM Nieminen, J. Phys.: Condens. Matter 11, 2017 (1999)., J. Phys. Condens. Matter. 11 (1999) 2017.

[14]  T. Gasi, A.K. Nayak, M. Nicklas, C. Felser, Structural and magnetic properties of the Heusler compound $Fe_2MnGa$, J. Appl. Phys. 113 (2013) 17E301.

[15]  T. Gasi, V. Ksenofontov, J. Kiss, S. Chadov, A.K. Nayak, M. Nicklas, J. Winterlik, M. Schwall, P. Klaer, P. Adler, others, Iron-based Heusler compounds $Fe_2YZ$: Comparison with theoretical predictions of the crystal structure and magnetic properties, Phys. Rev. B. 87 (2013) 64411.

[16]  M. Zhang, E. Brück, F.R. de Boer, G. Wu, Electronic structure, magnetism, and transport properties of the Heusler alloy $Fe_2CrAl$, J. Magn. Magn. Mater. 283 (2004) 409–414.

[17]  S. Kämmerer, A. Thomas, A. Hütten, G. Reiss, $Co_2MnSi$ Heusler alloy as magnetic electrodes in magnetic tunnel junctions, Appl. Phys. Lett. 85 (2004) 79–81.

[18]  S.A. Wolf, SA Wolf, DD Awschalom, RA Buhrman, JM Daughton, S. von Molnár, ML Roukes, AY Chtchelkanova, and DM Treger, Science 294, 1488 (2001)., Science (80-. ). 294 (2001) 1488.

[19]  R.A. De Groot, RA de Groot, FM Mueller, PG van Engen, and KHJ Buschow, Phys. Rev. Lett. 50, 2024 (1983)., Phys. Rev. Lett. 50 (1983) 2024.

[20]  Z. Ren, S.T. Li, H.Z. Luo, Structure and magnetic properties of Fe2CoGe synthesized by ball-milling, Phys. B Condens. Matter. 405 (2010) 2840–2843.

[21]  Y. Nishino, S. Deguchi, U. Mizutani, Thermal and transport properties of the Heusler-type Fe 2 VAl 1- x Ge x (0= x= 0.20) alloys: Effect of doping on lattice thermal conductivity, electrical resistivity, and Seebeck coefficient, Phys. Rev. B. 74 (2006) 115115.

[22]  Y. Nishino, Y. Nishino, M. Kato, S. Asano, K. Soda, M. Hayasaki, and U. Mizutani, Phys. Rev. Lett. 79, 1909 (1997)., Phys. Rev. Lett. 79 (1997) 1909.

[23]  S. V Dordevic, D.N. Basov, A. \ifmmode \acuteS\else Ś\filebarski, M.B. Maple, L. Degiorgi, Electronic structure and charge dynamics of the Heusler alloy ${\mathrm{Fe}}_{2}\mathrm{TiSn}$ probed by infrared and optical spectroscopy, Phys. Rev. B. 66 (2002) 75122.

[24]  B. Xu, L. Yi, Optical properties of the intermetallic compound Fe2TiSn, J. Phys. D. Appl. Phys. 41 (2008) 95404.

[25]  M. Mikami, S. Tanaka, K. Kobayashi, Thermoelectric properties of Sb-doped Heusler Fe2VAl alloy, J. Alloys Compd. 484 (2009) 444–448.

[26]  B. Xu, X. Li, G. Yu, J. Zhang, S. Ma, Y. Wang, L. Yi, The structural, elastic and thermoelectric properties of Fe2VAl at pressures, J. Alloys Compd. 565 (2013) 22–28.

[27]  H. Al-Yamani, B. Hamad, Thermoelectric properties of Fe 2 VAl and Fe 2 V 0.75 M 0.25 Al (M= Mo, Nb, Ta) alloys: first-principles calculations, J. Electron. Mater. 45 (2016) 1101–1114.

[28]  W. Lu, W. Zhang, L. Chen, Thermoelectric properties of (Fe1- xCox) 2VAl Heusler-type compounds, J. Alloys Compd. 484 (2009) 812–815.

[29]  M. Mikami, Y. Kinemuchi, K. Ozaki, Y. Terazawa, T. Takeuchi, Thermoelectric properties of tungsten-substituted Heusler Fe2VAl alloy, J. Appl. Phys. 111 (2012) 93710.

[30]  S. Sharma, S.K. Pandey, Investigation of the electronic and thermoelectric properties of Fe2ScX (X= P, As and Sb) full Heusler alloys by using first principles calculations, J.

Phys. D. Appl. Phys. 47 (2014) 445303.

[31] D.I. Bilc, G. Hautier, D. Waroquiers, G.-M. Rignanese, P. Ghosez, Low-dimensional transport and large thermoelectric power factors in bulk semiconductors by band engineering of highly directional electronic states, Phys. Rev. Lett. 114 (2015) 136601.

[32] H. Luo, G. Liu, F. Meng, J. Li, E. Liu, G. Wu, Half-metallicity in Fe-based Heusler alloys Fe2TiZ (Z = Ga, Ge, As, In, Sn and Sb), J. Magn. Magn. Mater. 324 (2012) 3295–3299. http://dx.doi.org/10.1016/j.jmmm.2012.05.033.

[33] S. Yabuuchi, M. Okamoto, A. Nishide, Y. Kurosaki, J. Hayakawa, Large Seebeck coefficients of Fe2TiSn and Fe2TiSi: First-principles study, Appl. Phys. Express. 6 (2013) 25504.

[34] I.H. Bhat, T.M. Bhat, D.C. Gupta, Magneto-electronic and thermoelectric properties of some Fe-based Heusler alloys, J. Phys. Chem. Solids. 119 (2018) 251–257.

[35] E. Haque, M.A. Hossain, First-principles study of elastic, electronic, thermodynamic, and thermoelectric transport properties of TaCoSn, Results Phys. 10 (2018) 458–465.

[36] N. Guechi, A. Bouhemadou, S. Bin-Omran, A. Bourzami, L. Louail, Elastic, Optoelectronic and Thermoelectric Properties of the Lead-Free Halide Semiconductors Cs2AgBiX6 (X= Cl, Br): Ab Initio Investigation, J. Electron. Mater. 47 (2018) 1533–1545.

[37] S. Sharma, B. Singh, P. Kumar, A comparative study of thermoelectric properties of CuGaTe2 by using PBE and MBJ potentials, in: AIP Conf. Proc., 2018: p. 140036.

[38] D.P. Rai, Sandeep, A. Shankar, R. Khenata, A.H. Reshak, C.E. Ekuma, R.K. Thapa, S.-H. Ke, Electronic, optical, and thermoelectric properties of Fe2+ x V1- x Al, AIP Adv. 7 (2017) 45118.

[39] D.J. Singh, Electronic structure calculations with the Tran-Blaha modified Becke-Johnson density functional, Phys. Rev. B. 82 (2010) 205102.

[40] E. Haque, M.A. Hossain, Origin of ultra-low lattice thermal conductivity in Cs2BiAgX6

(X= Cl, Br) and its impact on thermoelectric performance, J. Alloys Compd. 748 (2018) 63–72.

[41] M. Irfan, S. Azam, S. Hussain, S.A. Khan, M. Sohail, M. Ahmad, S. Goumri-Said, Enhanced thermoelectric properties of ASbO3 due to decreased band gap through modified becke johnson potential scheme, J. Phys. Chem. Solids. 119 (2018) 85–93.

[42] P. Hohenberg, W. Kohn, Inhomogeneous electron gas, Phys. Rev. 136 (1964) B864.

[43] P. Hohenberg, P. Hohenberg and W. Kohn, Phys. Rev. 136, B864 (1964)., Phys. Rev. 136 (1964) B864.

[44] M. Jamal, IRelast and 2DR-optimize packages are provided by M. Jamal as part of the commercial code WIEN2K, (2014).

[45] P. Blaha, K. Schwarz, G.K.H. Madsen, D. Kvasnicka, J. Luitz, wien2k, An Augment. Pl. Wave+ Local Orbitals Progr. Calc. Cryst. Prop. (2001).

[46] F. Tran, P. Blaha, Accurate band gaps of semiconductors and insulators with a semilocal exchange-correlation potential, Phys. Rev. Lett. 102 (2009) 226401.

[47] G.K.H. Madsen, D.J. Singh, BoltzTraP. A code for calculating band-structure dependent quantities, Comput. Phys. Commun. 175 (2006) 67–71.

[48] J.M. Ziman, @article{chaput2013direct, title={Direct solution to the linearized phonon Boltzmann equation}, author={Chaput, Laurent}, journal={Physical review letters}, volume={110}, number={26}, pages={265506}, year={2013}, publisher={APS} }, Oxford university press, 1960.

[49] C. Hurd, The Hall effect in metals and alloys, Springer Science & Business Media, 2012.

[50] G.B. Arfken, H.J. Weber, Mathematical methods for physicists international student edition, Academic press, 2005.

[51] A. Zakutayev, X. Zhang, A. Nagaraja, L. Yu, S. Lany, T.O. Mason, D.S. Ginley, A. Zunger, Theoretical prediction and experimental realization of new stable inorganic materials using the inverse design approach, J. Am. Chem. Soc. 135 (2013) 10048–10054.

[52]  S. Bhattacharya, G.K.H. Madsen, A novel p-type half-Heusler from high-throughput transport and defect calculations, J. Mater. Chem. C. 4 (2016) 11261–11268.

[53]  Gautier R., Zhang X., Hu L., Yu L. and Zunger A., Nat. Chem., 7 (2015) 308.

[54]  H. Ehrenreich, H. Ehrenreich and MH Cohen, Phys. Rev. 115, 786 (1959), Phys. Rev. 115 (1959) 786.

[55]  C.C. Kim, CC Kim, JW Garland, H. Abad, and PM Raccah, Phys. Rev. B 45, 11749 (1992)., Phys. Rev. B. 45 (1992) 11749.

[56]  H.A. Kramers, La diffusion de la lumiere par les atomes, 1927.

[57]  R. de L. Kronig, On the theory of dispersion of x-rays, Josa. 12 (1926) 547–557.

[58]  D.E. Aspnes, DE Aspnes and AA Studna, Phys. Rev. B 27, 985 (1983)., Phys. Rev. B. 27 (1983) 985.

[59]  R.E. LaVilla, H. Mendlowitz, Optical Properties of Germanium, J. Appl. Phys. 40 (1969) 3297–3300.

[60]  J. Tauc, R. Grigorovici, A. Vancu, Optical properties and electronic structure of amorphous germanium, Phys. Status Solidi. 15 (1966) 627–637.

[61]  G. Dolling, G. Dolling and RA Cowley, Proc. Phys. Soc.(London) 88, 463 (1966), in: Proc. Phys. Soc. London, 1966: p. 463.

[62]  J.B. Theeten, D.E. Aspnes, R.P.H. Chang, A new resonant ellipsometric technique for characterizing the interface between GaAs and its plasma-grown oxide, J. Appl. Phys. 49 (1978) 6097–6102.

[63]  J.S. Blakemore, Semiconducting and other major properties of gallium arsenide, J. Appl. Phys. 53 (1982) R123--R181.

[64]  M. Sturge, Optical absorption of gallium arsenide between 0.6 and 2.75 eV, Phys. Rev. 127 (1962) 768.

[65]  J.O. Akinlami, A.O. Ashamu, Optical properties of GaAs, J. Semicond. 34 (2013) 32002.

[66]   K. Bücher, J. Bruns, H.G. Wagemann, Absorption coefficient of silicon: An assessment of measurements and the simulation of temperature variation, J. Appl. Phys. 75 (1994) 1127–1132.

[67]   C.M.I. Okoye, Optical properties of the antiperovskite superconductor MgCNi3, J. Phys. Condens. Matter. 15 (2003) 833.

[68]   G.J. Snyder, E.S. Toberer, Complex thermoelectric materials, in: Mater. Sustain. Energy A Collect. Peer-Reviewed Res. Rev. Artic. from Nat. Publ. Gr., World Scientific, 2011: pp. 101–110.

[69]   P. Bujard, E. Walker, Elastic constants of Cr3Si, Solid State Commun. 39 (1981) 667–669.

[70]   A.. Dulong, P.L. and Petit, Recherches sur quelques points important de la thrie de la chaleur, Ann. Chim. Phys. 10 (1819) 395–413.

[71]   K. Kaur, TiPdSn: A half Heusler compound with high thermoelectric performance, EPL (Europhysics Lett. 117 (2017) 47002.

Figures

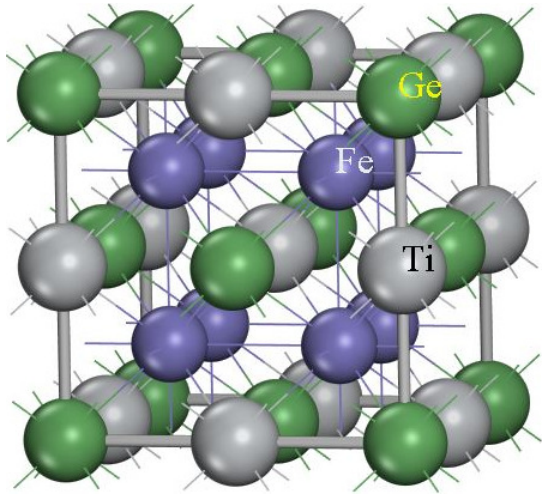

**Fig. 1: E**quilibrium crystal structure of Fe$_2$TiGe.

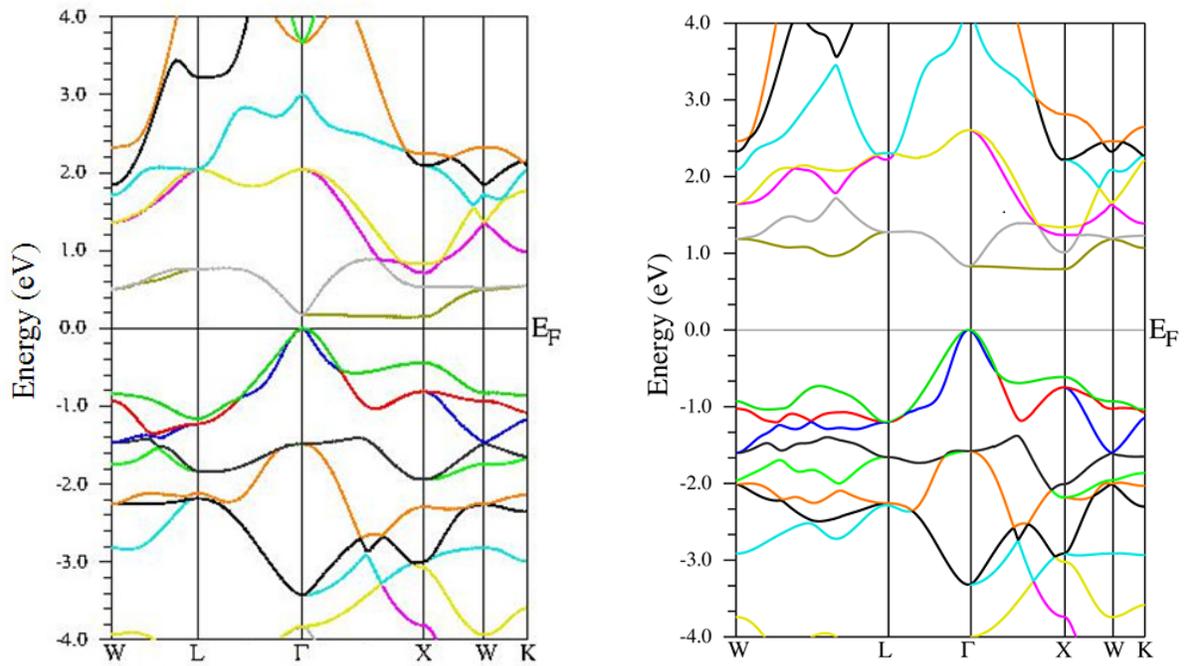

Fig. 2. Electronic band structure of Fe$_2$TiGe using: (a) PBE-GGA and (b) TB-mBJ density functional.

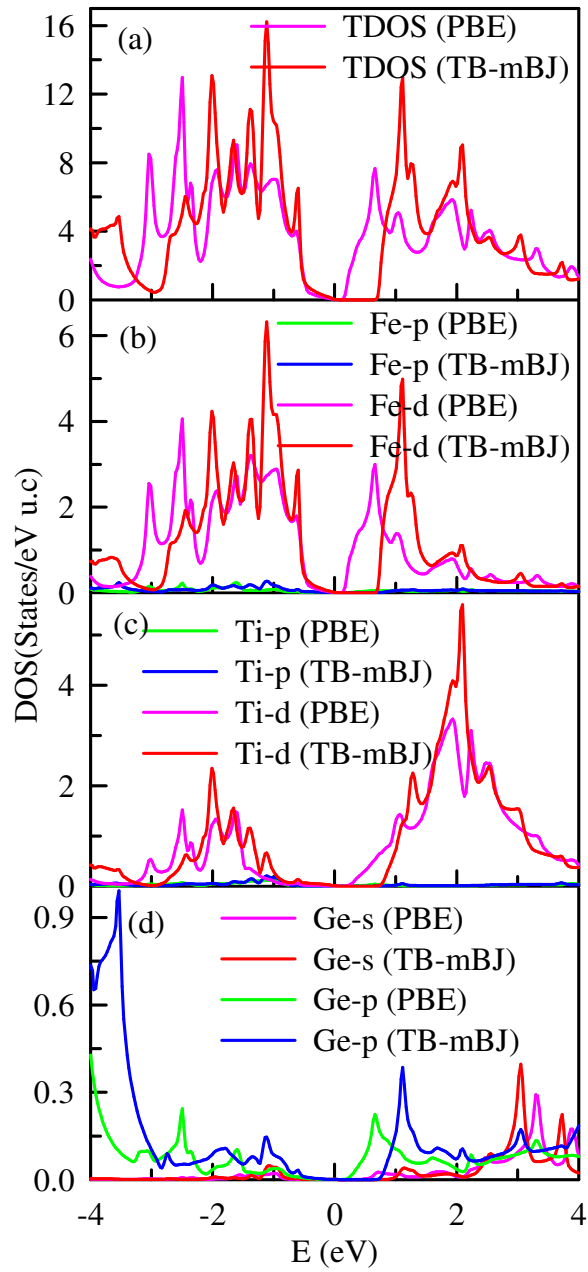

Fig. 3. Total and projected density of states of $Fe_2TiGe$ using: (a) PBE-GGA and (b) TB-mBJ density functional.

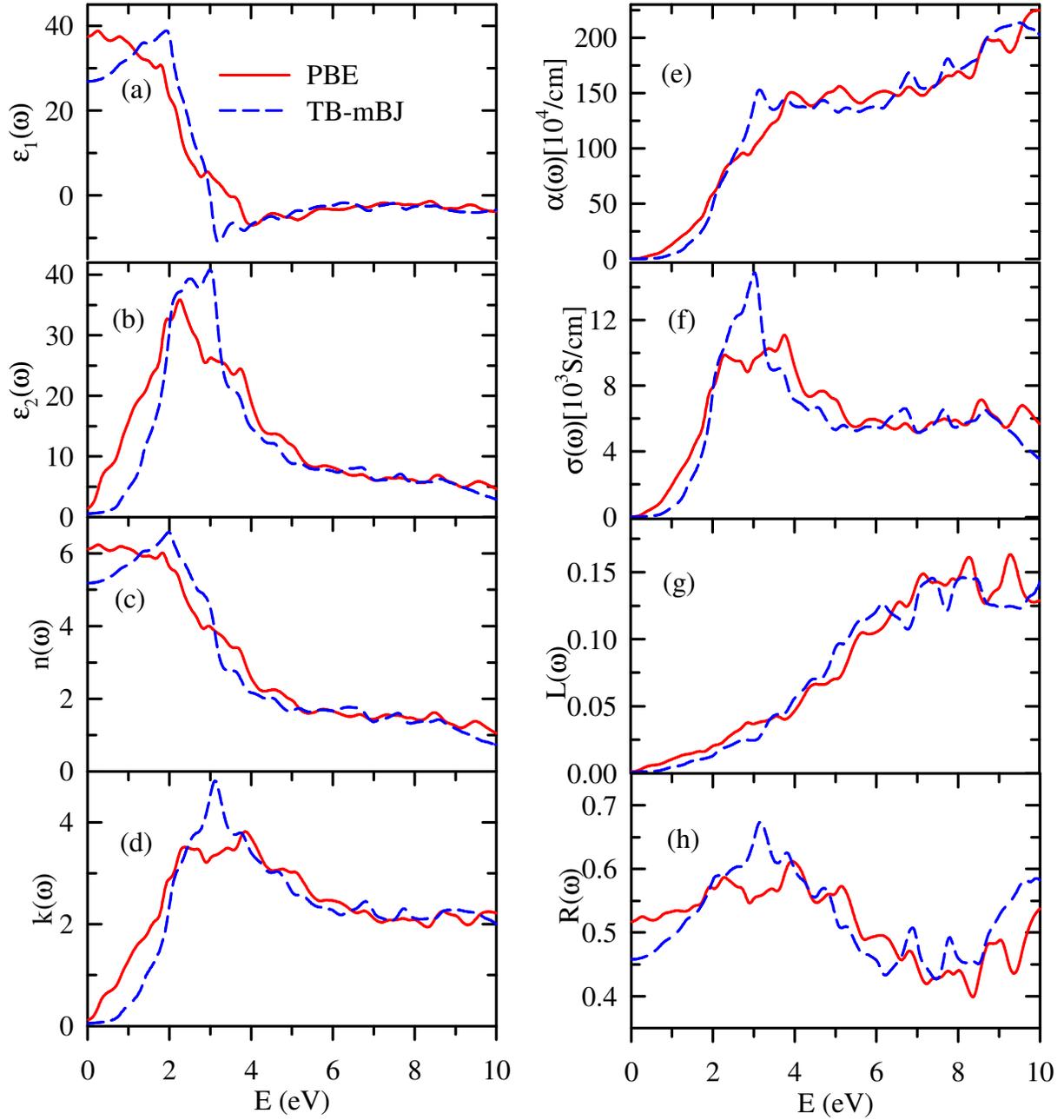

Fig. 4. Optical properties of Fe$_2$TiGe: (a) real dielectric function $\varepsilon_1(\omega)$, (b) imaginary dielectric function $\varepsilon_2(\omega)$, (c) refractive index (n), (d) extinction coefficient (k), (e) absorption coefficient ($\alpha$), (f) optical conductivity ($\sigma$), (g) loss function and (h) reflectivity (R).

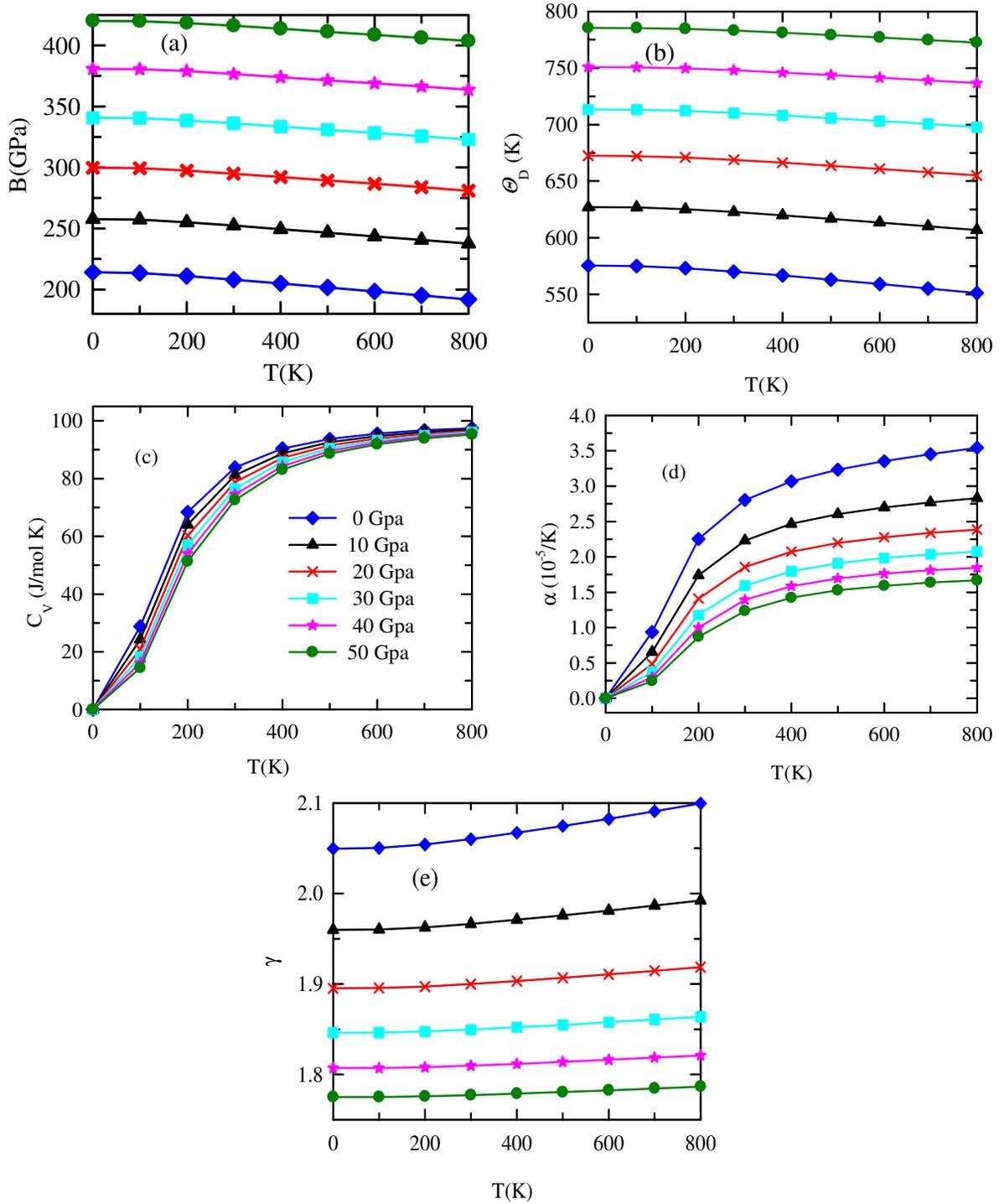

Fig. 5. Temperature and pressure dependent thermodynamic properties of $Fe_2TiGe$: (a) bulk modulus, (b) Debye temperature, (c) specific heat at constant volume, (d) volume thermal expansion coefficient and (e) Gruneisen parameter.

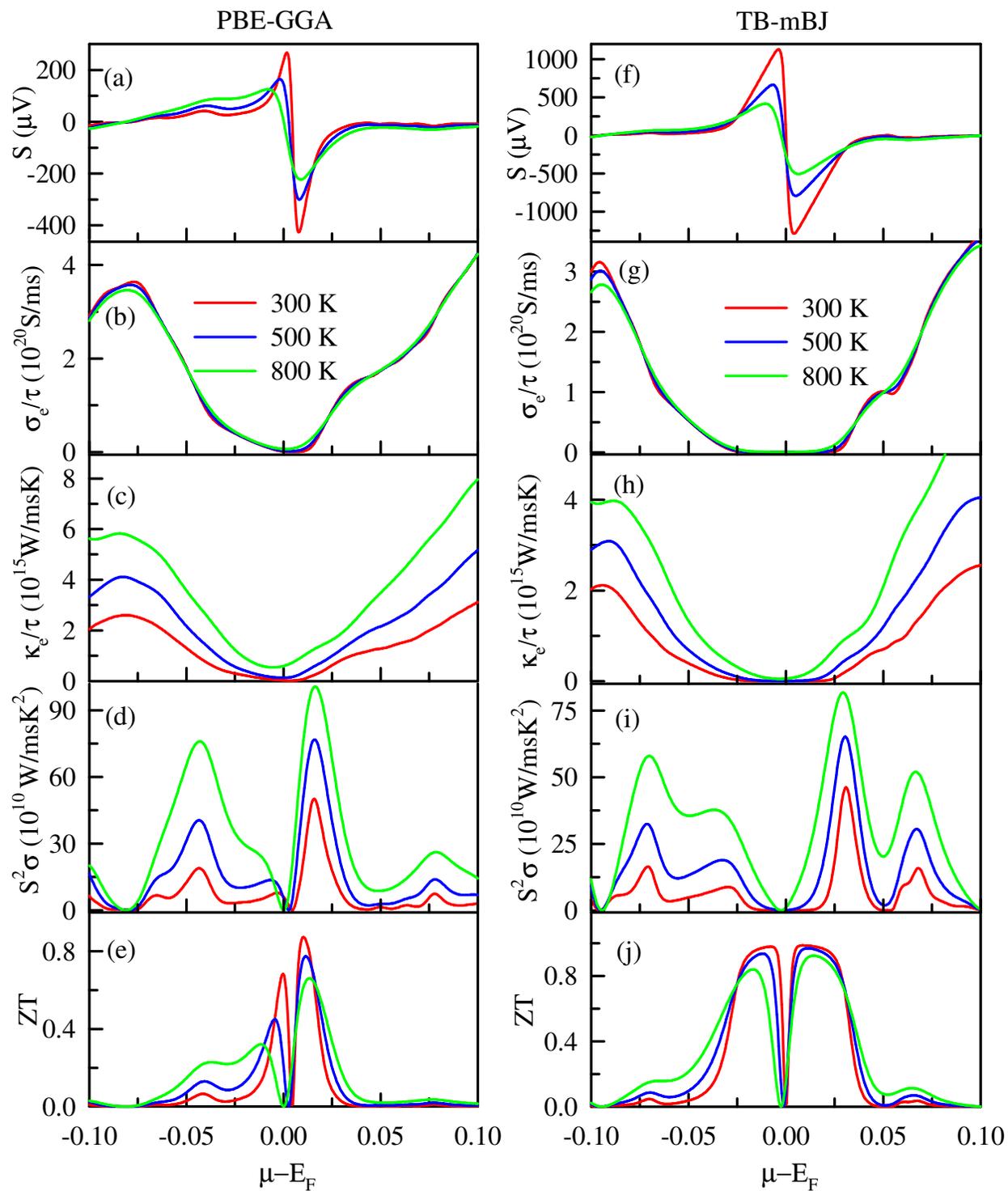

Fig. 6. Chemical potential (μ) dependent thermoelectric parameters using PBE-GGA (left panel) and TB-mBJ potentials (right panel).

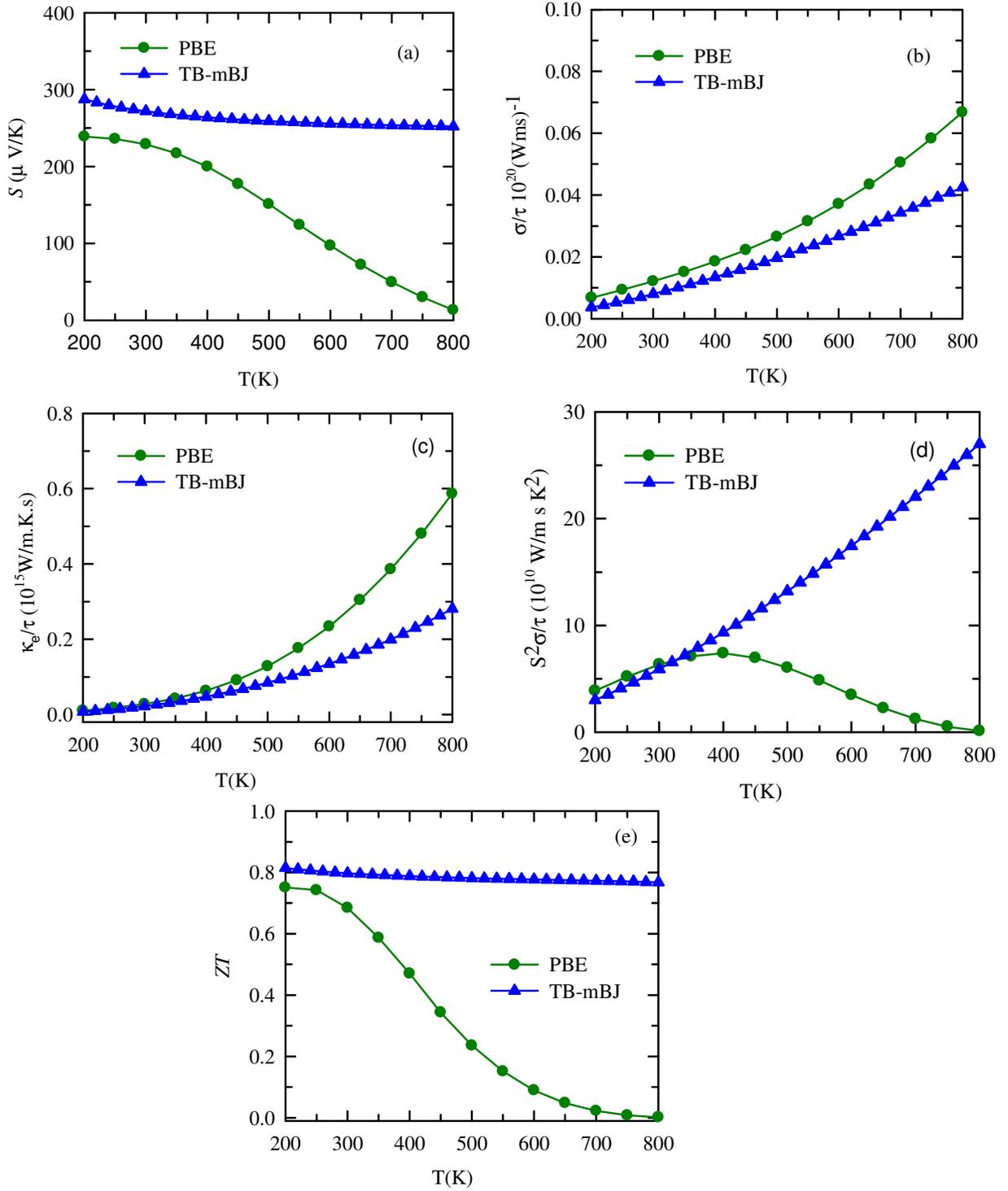

Fig. 7. Temperature dependence of thermoelectric transport properties of $Fe_2TiGe$: (a) Seebeck Coefficient, (b) Electrical conductivity, (c) Electronic thermal conductivity, (d) Power factor, (e) *ZT*.